\documentclass[11pt,nofootinbib,showpacs]{revtex4}
\usepackage{amsfonts,amssymb,amsmath,graphicx}
\def\dac{\displaystyle\frac}
\def\[{\left[}
\def\]{\right]}
\def\({\left(}
\def\){\right)}
\def\gammad{\gamma _{\left( \mathbf{D}\right)}}
\def\gammaterm{\gamma _{\left( \mathbf{D}\right)}+\dot b(t)^2}
\def\bb{b(t)}
\def\db{\dot b(t)}
\def\ddb{\ddot b(t)}
\def\hh{H(t)}
\def\dh{\dot H(t)}

\newcommand{\const}{\mathop{\rm const}\nolimits}

\begin{document}
\baselineskip7mm

\title{Cosmological dynamics in higher-dimensional Einstein-Gauss-Bonnet gravity}

\author{Fabrizio Canfora}
\affiliation{Centro de Estudios Cientificos (CECs), Casilla 1469 Valdivia, Chile}
\affiliation{Universidad Andres Bello, Av. Republica 440, Santiago, Chile}
\author{Alex Giacomini}
\affiliation{Instituto de Ciencias F\'isicas y Matem\'aticas, Universidad Austral de Chile, Valdivia, Chile}
\author{Sergey A. Pavluchenko}
\affiliation{Instituto de Ciencias F\'isicas y Matem\'aticas, Universidad Austral de Chile, Valdivia, Chile}

\begin{abstract}
In this paper we perform a systematic classification of the regimes of cosmological dynamics in  Einstein-Gauss-Bonnet gravity with generic values of the
coupling constants.  We consider a manifold   which is a warped product of a four dimensional Friedmann-Robertson-Walker space-time   with a $D$-dimensional
Euclidean compact constant curvature space with two independent scale factors.
A numerical analysis of the time evolution as function of the coupling constants and of the curvatures of the spatial section and
of the extra dimension is performed. We describe the distribution of the regimes over the initial conditions space and the coupling constants. The analysis
is performed for two
values of the number of extra dimensions ($D\geqslant 6$ both) which allows us to describe the effect of the number of the extra dimensions as well.
\end{abstract}

\pacs{04.50.-h, 11.25.Mj, 98.80.-k}

\maketitle

\section{Introduction}
The action principle for General Relativity in four dimensions can be formulated by requiring that  it should be constructed out of curvature invariants and
that its variation  should lead to second order field equations for the metric. Indeed in four dimensional space-time it can be easily shown that the only
action that satisfies these two requirements is the Einstein-Hilbert action plus a cosmological term. Rigorously speaking one can add to the gravitational
action also a Gauss-Bonnet term $\int \sqrt{-g}(R_{\mu\nu\alpha\beta}R^{\mu\nu\alpha\beta}-4R_{\mu\nu}R^{\mu\nu}+R^2)$ as its variation does not affect the
equations of motion being just a boundary term. This is  only true in four dimensions as in five or higher dimensions the variation of the Gauss-Bonnet term
gives a non-trivial contribution to the equations of motion which however remains of second order in the derivatives of the metric. In seven dimensions one
can add another  term to the gravitational action which is constructed from  cubic curvature invariants and whose variation is again of second order in the
metric. More generally in an $N$-dimensional space-time it is possible to construct an action which is the sum of  $[\frac{N+1}{2}]$ independent terms which
are of up to the $[\frac{N}{2}]$th power in the curvature and whose variation gives second order field equations for the metric. This gravity theory is
known as Lovelock gravity~\cite{Lovelock}, and it is the most straightforward extension of General Relativity to higher dimensions as it is
constructed according to the same principles as used for four dimensional General Relativity.

Einstein-Gauss-Bonnet (EGB) gravity is the most simple non-trivial example of Lovelock gravity and its action is the sum of a volume term $\int \sqrt{-g}$ an
Einstein-Hilbert term $\int \sqrt{-g}R$and the Gauss-Bonnet term $\int \sqrt{-g}(R_{\mu\nu\alpha\beta}R^{\mu\nu\alpha\beta}-4R_{\mu\nu}R^{\mu\nu}+R^2)$,
each term having its own coupling constant. The reason why this action has attracted a lot of interest in theoretical physics is also because  the low
energy  limit of some string theories (see, for instance, the discussion in~\cite{GastGarr})  is described  by  Einstein-Gauss-Bonnet gravity
rather then General Relativity. Actually the idea that space-time may have more than four dimensions is much older than string theory as it was already
proposed by Kaluza~\cite{KK1} and \mbox{Klein~\cite{KK2, KK3}} in an attempt to unify gravity with the
electromagnetic interaction. Originally one extra dimension was considered. In order to encompass also non-Abelian gauge fields, more extra dimensions must
be added.

Assuming the existence of extra dimensions arises of course the question of why these are not visible to us.
The most standard explanation of this is to assume that the extra dimensions are compactified to a very small scale,
 which however opens the question of why the extra dimensions are of almost constant size in time while the three macroscopic space  dimensions are 
 expanding at approximately constant rate. In \cite{CGTW}, it was shown for the first time  how to construct a
realistic static compactification  in seven (or higher) dimensions in Lovelock
gravities. A suitable class of cubic Lovelock theories allows to recover
General Relativity with an arbitrarily small positive cosmological constant in four dimensions
and with the extra dimensions of constant curvature. The drawback of this construction is that it involves a fine tuning between the coupling constants and 
moreover it does not provide a dynamical description of the compactification. This is actually of great interest as it may be that in the far past the extra 
dimensions were of approximately the same size as the three macroscopic space dimensions.

Historically compactification of Lovelock gravity  has begun to attract the interest of researchers already in the 80s~\cite{add_1, add_2, add_3, add_4, Is86, 44, add_5, add_6, 
add_9, add_10, add_7, add_8}.  
The first question that arises in this context is the existence exact or static compactified solutions where the metric is
a cross product of a 3+1 dimensional manifold times a constant curvature ``inner space''. Such an exact static solution can 
be interpreted as a ground state of the Kaluza-Klein theory. These exact solutions are also known as ``spontaneous 
compactification'' in literature. The existence of such solutions  were first discussed in~\cite{add_1} where the four dimensional 
Lorentzian factor was actually Minkowski (the generalization for a constant curvature Lorentzian mainifold was done in~\cite{add_9}).

In the context of a cosmological model it is necessary to study a metric with time dependent scale factors. The most simple ansatz is given by 
assuming a constant size of the extra dimensions and the four dimensional factor being a FRW manifold. This has been studied in~\cite{add_4} where the 
attention was put on exact solutions with exponential scale factor. In the last paper it was explicitly stated that a more realistic model needs
to consider the dynamical evolution of the extra dimensional scale factor. In the context of exact solutions such an attempt was done in~\cite{Is86}
where both the 3-space and the extra dimensional scale factors where exponential functions. Solutions with exponentially increasing 3-D scale 
factor and exponentially shrinking extra dimensional scale factor were found.
The drawback of these solutions was that an exponentially shrinking to zero radius of the extra dimensions is phenomenologically problematic as 
the radius of the extra dimensions is related to the strength of the gauge field  couplings. 

In~\cite{add_9} the structure of the equations of motion for 
Lovelock theories for various types of solutions has been studied. It was stressed that the Lambda term in the action is actually not a 
cosmological constant as it does not give the curvature scale of a maximally symmetric manifold. In the same paper the equations of motion for 
compactification with both time dependent scale factors were written for arbitrary Lovelock order in the special case that both factors are flat. 
The results of~\cite{add_9} were reanalized for the special case of 10 space-time dimensions in~\cite{add_10}.
In~\cite{add_8} the existence of dynamical compactification solutions was studied with the use of Hamiltonian formalism.

More recent analysis focuses on
properties of black holes in Gauss-Bonnet~\cite{addn_1, addn_2} and Lovelock~\cite{addn_3, addn_4} gravities, features of gravitational collapse in these
theories~\cite{addn_5, addn_6, addn_7}, general features of spherical-symmetric solutions~\cite{addn_8} and many others. Cosmological ``counterpart'' of
this field was also intensively studied both numerically for a wide variety of cosmological models~\cite{add13, add_12, mpla09, grg10, gc10, prd10, CGP1, MO04, MO14} and
analitically mostly in attempts to find exact solutions~\cite{iv10-1, iv10-2, prd09, grg10, gc10, new12, new13, new14}. Of particular relevance are \cite{add13} where
the dynamical compactification of (5+1) EGB model was considered, \cite{MO04, MO14}, with the metric {\it ansatz} different from what we are about to consider,
and \cite{CGP1} of which the present paper is direct continuation in the sense that now we consider all possible regimes. To be more precise, 

In all the cited papers the search for  solutions never went beyond the ansatz of exponential scale factor or beyond the search of criteria for 
the existence of solutions. This is due to the fact that the equations of motion are too difficult to find an exact generic solution even in the 
case of constant size of the extra dimensions. On the other hand it is of great interest to analyze more generic solutions in order to understand 
all the possible regimes of dynamic compactification. In order to do this a numerical analysis must be performed. This has been done in~\cite{MO14} where
the scale factor of the extra dimensions again was kept constant. It is however of great interst to analyze the problem also with a dynamical 
scale factor for the extra dimensions as a constant extra dimensional scale factor would imply a strong conspiracy of the initial conditions whose
space may in this case have zero measure. The aim of this paper is therefore to make a numerical analysis for dynamical compactification where 
both scale factors are time dependent and where no a priori assumptions are made on the functional form of the scale factor or on the sign of the 
curvature of both factors of space-time.

Another aim of this paper is to get a systematic classification all possible dynamical compactification regimes  of the most generic Einstein-Gauss-Bonnet
gravity in arbitrary dimensions and with arbitrary coupling constants. In order to perform this analysis we will make an {\it ansatz} of a  warped product
space-time of the form \mbox{$M_4\times b(t)M_D$}, where $M_4$ is a Friedmann-Robertson-Walker manifold with scale factor $a(t)$ whereas
$M_D$ is a $D$-dimensional Euclidean compact and constant curvature manifold with scale factor $b(t)$. We will then study the dynamical evolution of the two
scale factors. To be as generic as possible curvature will considered in both in the extra dimensions and in the spatial section of the four dimensional
part of the metric. This is of special interest as in most cases, when considering  Gauss-Bonnet, or even higher-order Lovelock gravity, in literature  only
spatially flat sections are considered. Considering the case with non-zero spatial constant curvature  allows to
see the influence of the curvature on the cosmological dynamics. Despite the fact that, according to current observational cosmological data, our Universe
is flat with a high precision, at the early
stages of the Universe evolution the curvature could comes into play. In there, negative curvature only ``helps'' inflation (since the effective equation of
state for curvature is $w=-1/3$),
while the positive curvature affects the inflationary asymptotics, but its influence is not strong for a wide variety of the scalar field potentials
(see~\cite{infl1, infl2} for details), so that
we can safely consider both signs for curvature without worrying for inflationary asymptotics.
The equations of motion are  highly nonlinear and therefore it is technically impossible to integrate them in a closed form.
However it is possible to understand in detail all the relevant features of the theory, depending on the values of the couplings
and of the curvature of space and extra dimension, by performing a numerical analysis.

In order to make a complete classification of all possible regimes of the dynamical compactification it is important to notice that the volume term in in
Einstein-Gauss-Bonnet gravity, is not directly related to the cosmological constant. To demonstrate it let us consider EGB action in the vielbein formalism

\begin{equation}
S=\int \epsilon_{{A_1} \ldots A_{D+4}}(\frac{c_0}{D+4}e^{A_1}\ldots e^{A_{D+4}}+\frac{c_1}{D+2}R^{{A_1}{A_2}}e^{A_3}\ldots e^{A_{D+4}}+\frac{c_2}{D}R^{A_1 A_2}R^{A_3 A_4}e^{A_5}\ldots e^{A_{D+4}}),
\label{actionEGB}
\end{equation}

\noindent and vary it with respect to the vielbein to obtain equations of motion

\begin{equation}
\epsilon_{A_1}=\epsilon_{{A_1} \ldots A_{D+4}}({c_0}e^{A_2}\ldots e^{A_{D+4}}+{c_1}R^{{A_2}{A_3}}e^{A_4}\ldots e^{A_{D+4}}+{c_2}R^{A_2 A_3}R^{A_4 A_5}e^{A_6}\ldots e^{A_{D+4}})\label{EOMD+4}.
\end{equation}

The ``cosmological constant'' indeed measures the curvature scale of a maximally-symmetric space-time solution. A maximally symmetric space-time has  curvature two form given by

\begin{equation}
R^{AB}=\Lambda_{D+4}e^{A}e^{B}\label{constantcurv},
\end{equation}

\noindent which inserted in the equations of motion gives a quadratic equation for $\Lambda_{D+4}$:

\begin{equation}
(c_2\Lambda_{D+4}^2 + c_1\Lambda_{D+4}+c_0)e^{A_1}\ldots e^{A_{D+4}}=0\label{polynomial};
\end{equation}

\noindent which admits as solutions

\begin{equation}
\Lambda_{eff}\equiv\Lambda_{D+4}= \frac{-c_1\pm \sqrt{(c_1)^2-4c_2c_0}}{2c_2}\label{Leff}.
\end{equation}

Due to the fact that the Einstein-Gauss-Bonnet action is quadratic in the curvature, the equations of motion for a
maximally symmetric space-time {\it ansatz} will give a quadratic equation for the curvature scale (\ref{polynomial}). This means that, in general,
Einstein-Gauss-Bonnet gravity admits up to two maximally symmetric space-time solutions (\ref{Leff}). It is however also possible that the discriminant of
the quadratic equation (\ref{polynomial}) is negative
so that a maximally symmetric solution does not exist at all. This special situation in combination with a negative curvature of the extra dimensions has
been studied in \cite{CGP1} and it was the first time where a phenomenologically reasonable dynamical compactification scenario, where the size of the extra dimensions
and the four dimensional Hubble parameter tend to a constant, without fine tunings or ad hoc matter fields was found.

Another important feature of this theory, in opposition to GR, is that by compactifying the space-time to $M_4\times M_D$ where $M_4$ is a four dimensional space-time and $M_D$ is some compact manifold with
constant curvature $\Lambda_D$ is that the Newton constant of the effective four dimensional theory is not just proportional to $c_1$. This can be seen by projecting the $4+D$ dimensional equations down to four dimensions

\begin{equation}
\begin{array}{l}
\epsilon_{i}=\(c_1(D+1)+2c_2(D-1)\Lambda_D\)R^{jk}+\(c_0\frac{(D+3)(D+2)(D+1)}{6}+\right. \\ \left.
+c_1\Lambda_D\frac{(D+1)D(D-1)}{6}+c_2\Lambda_D^2\frac{(D-1)(D-2)(D-3)}{6}\)e^ie^j=0\label{effectivenewton}.
\end{array}
\end{equation}
where the lowercase indices run from zero to three.
The term $(c_1(D+1)+2c_2(D-1)\Lambda_D)$ which multiplies the four dimensional curvature two form $R^{jk}$ is the ``effective Newton constant'' whereas
the term $(c_0\frac{(D+3)(D+2)(D+1)}{6}+c_1\Lambda_D\frac{(D+1)D(D-1)}{6}+c_2\Lambda_D^2\frac{(D-1)(D-2)(D-3)}{6})$ is an ``effective 4-dimensional cosmological constant''.
This means that if the Gauss-Bonnet term does not vanish and moreover the $D$-dimensional curvature does not vanish the effective Newton constant is not just proportional to $c_1$. In particular, the
effective Newton constant  can even have a negative sign.

In this paper we want to obtain a more complete picture by studying all the possible dynamical compactification regimes for the most generic
Einstein-Gauss-Bonnet gravity. A numerical analysis making a scan over a reasonably broad region of the initial conditions space will be performed. Special
attention will be given to the late time evolution of the scale factors and if they describe a phenomenologically sensible scenario.

The structure of the paper will be the following: in the next section the detailed numerical analysis and some basic discussion are performed and in
the last section we discuss the results in detail and summarize them.

\section{Numerical analysis}

The {\it ansatz} for the metric is
\begin{equation}
ds^{2}=-dt^{2}+a(t)^{2}d\Sigma _{(3)}^{2}+b(t)^{2}d\Sigma _{(\mathbf{D}%
)}^{2}\ ,  \label{Ansatz-metric}
\end{equation}%

\noindent where $d\Sigma _{(3)}^{2}$ and $d\Sigma _{(\mathbf{D})}^{2}$ stand for the
metric of two constant curvature manifolds $\Sigma _{(3)}$ and $\Sigma
_{(\mathbf{D})}$. It is worth to point out that even a negative constant curvature space can be compactified by making the quotient of the space by a
freely acting discrete subgroup of $O(D,1)$ \cite{wolf}.

The complete derivation of the equations of motion could be found in our previous paper, dedicated to the description of the particular regime which appears in this model~\cite{CGP1}.
In the first part of the analysis we will for simplicity  consider the case with $\gamma _{\left( 3\right)} = 0$ (zero spatial curvature for ``our'' (3+1)-dimensional world; we will then  later, in order to be as generic as possible also analyze effect of a non-zero $\gamma_3$ on the compactification regimes. For the moment the non-zero curvature for extra dimensions can be normalized as
$\gamma _{\left( \mathbf{D}\right)} = \pm 1$. Since there is no curvature
term for $a(t)$, it is useful to rewrite the equations of motion in terms of the Hubble parameter $H(t) = \dot a(t)/a(t)$; the equations will take a form

%\begin{equation}
%\begin{array}{l}
%\end{array}
%\end{equation}

\small
\begin{equation}
\begin{array}{l}
\dac{(D+3)(D+2)(D+1)}{6}c_0 + \dac{(D+1)(D-1)}{6}c_1 \[ \dac{\gammaterm}{\bb^2} + \dac{6\hh\db}{\bb(D-1)} + \dac{6\hh^2}{D(D-1)}  \] + \\ \\ +
\dac{(D-1)(D-2)(D-3)}{6}c_2 \[ \dac{(\gammaterm)^2}{\bb^4} + \dac{12\hh^2(\gammaterm)}{\bb^2(D-2)(D-3)} + \dac{24\hh^2\db^2}{\bb^2(D-2)(D-3)} + \right. \\ \\ \left. +
\dac{12(\gammaterm)\hh\db}{\bb^3(D-3)} + \dac{24\hh^3\db}{\bb(D-1)(D-2)(D-3)} \]=0,
\end{array}\label{eq_my_1}
\end{equation}
%\normalsize

\begin{equation}
\begin{array}{l}
\dac{(D+3)(D+2)(D+1)}{6}c_0 + \dac{(D+1)(D-1)}{6}c_1 \[ \dac{\gammaterm}{\bb^2} + \dac{4(\dh+\hh^2)}{D(D-1)} + \dac{2\ddb}{\bb(D-1)} + \right. \\ \\ \left. +\dac{2\hh^2}{D(D-1)} +
\dac{4\hh\db}{\bb(D-1)}  \] + \dac{(D-1)(D-2)(D-3)}{6}c_2 \[ \dac{(\gammaterm)^2}{\bb^4} + \right. \\ \\ \left. + \dac{16(\dh+\hh^2)\hh\db}{\bb(D-1)(D-2)(D-3)} +
\dac{8(\gammaterm)\hh\db}{\bb^3(D-3)} + \dac{8(\dh+\hh^2)(\gammaterm)}{\bb^2(D-2)(D-3)} + \right. \\ \\ \left. + \dac{8\hh^2\ddb}{\bb(D-1)(D-2)(D-3)} +
\dac{16\ddb\hh\db}{\bb^2(D-2)(D-3)} + \dac{4\ddb(\gammaterm)}{\bb^3(D-3)} + \right. \\ \\ \left. + \dac{4\hh^2 (\gammaterm)}{\bb^2(D-2)(D-3)} + \dac{8\hh^2\db^2}{\bb^2(D-2)(D-3)}  \]=0,
\end{array}\label{eq_my_2}
\end{equation}
%\normalsize

\begin{equation}
\begin{array}{l}
\dac{D(D+3)(D+2)(D+1)}{6}c_0 + \dac{(D-2)(D+1)(D-1)}{6}c_1 \[ \dac{\gammaterm}{\bb^2} + \dac{6(\dh+\hh^2)}{(D-2)(D-1)} + \right. \\ \\ \left. \dac{2\ddb}{\bb(D-2)} +
\dac{6\hh^2}{(D-2)(D-1)} +
\dac{6\hh\db}{\bb(D-2)}  \] + \dac{(D-1)(D-2)(D-3)}{6}c_2 \times \\ \\ \times \[ \dac{(\gammaterm)^2}{\bb^4}(D-4) + \dac{48(\dh+\hh^2)\hh\db}{\bb(D-2)(D-3)} +
\dac{12(\gammaterm)\hh\db}{\bb^3} + \right. \\ \\ \left. + \dac{24\hh^2\db^2}{\bb^2(D-3)} + \dac{12(\dh+\hh^2)(\gammaterm)}{\bb^2(D-3)} + \dac{24\hh^2\ddb}{\bb(D-2)(D-3)} +
\dac{24\ddb\hh\db}{\bb^2(D-3)} + \right. \\ \\ \left. + \dac{4\ddb(\gammaterm)}{\bb^3} + \dac{12\hh^2 (\gammaterm)}{\bb^2(D-3)} + \dac{24\hh^3\db}{\bb(D-2)(D-3)} +
\dac{24(\dh+\hh^2)\hh^2}{(D-1)(D-2)(D-3)} \]=0,
\end{array}\label{eq_my_3}
\end{equation}
\normalsize

\noindent as $\mathcal{E}_{0}=0$ equation (\ref{eq_my_1}), $\mathcal{E}_{i}=0$ (\ref{eq_my_2}), and  $\mathcal{E}_{a}=0$ (\ref{eq_my_3}). One can see from the system that it is too
complicated to look for some analytic solutions so we have to analyze it numerically. Still, we have three initial conditions ($H_0 = H|_{t=0}$, $b_0 = b|_{t=0}$ and $\dot b_0 =
\dot b|_{t=0}$) and five parameters ($c_0$, $c_1$, $c_2$, $D$, and $\gammad$); initial conditions are bound by the constraint equation (\ref{eq_my_1}) so that there are only two independent
initial conditions. For such we chose $H_0$ and $b_0$ -- both should be positive and we want to be sure of it by putting them so by hand. So the procedure is as follows -- we
set the values for parameters, fix some initial $H_0$ and $b_0$, calculate $\dot b_0$ from (\ref{eq_my_1}) and, by solving (\ref{eq_my_2}) and (\ref{eq_my_3}) numerically, calculate
the evolution both forward and back in time to see the whole evolution of the cosmological model.
We repeat the procedure to make a
scan over $H_0$ and $b_0$ within the reasonable range. The ``reasonable'' here requires explanation -- since we work with
both Einstein-Hilbert and Gauss-Bonnet contributions, we have terms that are linear and squared on curvature. For that we want to cover the energy scales which allow both terms to
dominate. Say, with the coupling constants of the order of unity the range for $H_0$ and $b_0$ from decimals to dozens would cover the initial dominance of both terms.

\begin{figure}
\includegraphics[width=1.0\textwidth, angle=0]{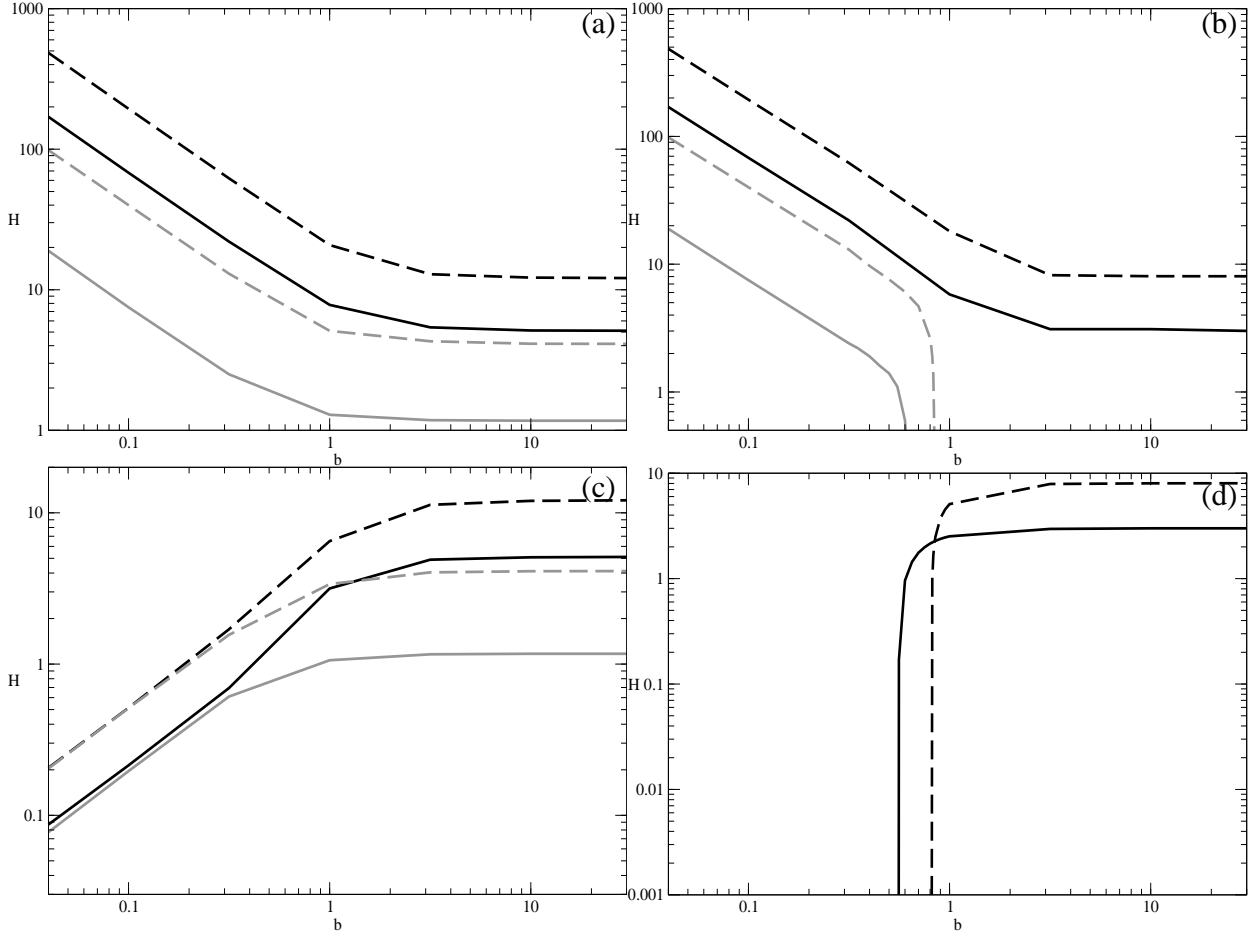}
\caption{Distribution of the regions with different number of roots over the initial conditions space ($b\equiv b_0 = b|_{t=0}$ and $H\equiv H_0 = H|_{t=0}$) for the cases
$\gammad=+1,\,\{c_0, c_1, c_2\} = \{1, 1, 1\}$ in (a) panel, $\gammad=+1,\,\{c_0, c_1, c_2\} = \{-1, 1, 1\}$ in (b) panel, $\gammad=-1,\,\{c_0, c_1, c_2\} = \{1, 1, 1\}$ in (c)
panel, and $\gammad=-1,\,\{c_0, c_1, c_2\} = \{-1, 1, 1\}$ in (d) panel. Black line separate regions with four root (above the line) from region with two roots (below); grey line
separate the region with two roots (above) from the region with no roots (below). Solid line corresponds to the $D=6$ case, while dashed -- to the $D=16$ case
(see text for details).}\label{fig_distr}
\end{figure}

%It appears that all possible regime distributions could be brought to four cases depending on the effective cosmological constant (see below). Two cases correspond to $\gammad=+1$,
%they are presented in Fig. \ref{fig_distr} (a) and (b) panels.

In Fig. \ref{fig_distr} we plotted the distribution of the regions with different number of roots over the initial conditions space
($b\equiv b_0$ and $H\equiv H_0$). The roots in question are the roots of the eq. (\ref{eq_my_1}) -- one can see that it is fourth order with respect to $\dot b_0$,
so up to four real roots exists and the regimes and their distribution depends both on the number of roots and which branch (root) we choose. Here we explain and comment the roots
distribution and below we give details on the regimes themselves. So in Fig. \ref{fig_distr} we plotted these distribution: black line separate regions with four root (above the
line) from region with two roots (below); grey line separate the region with two roots (above) from the region with no roots (below). Solid line corresponds to the $D=6$ case,
while dashed -- to the $D=16$ case -- we plotted them both on the same graph to demonstrate the effect of the number of extra dimensions on the distribution.

Despite the fact that there is a continuous distribution of the three coupling constants ($c_0$, $c_1$, and $c_2$), all the cases for given $\gammad$ could 
be brought to two  -- with positive and negative discriminant of (\ref{polynomial}) -- the qualitative picture remains the same for the same $\gammad$ and
all combinations of ($c_0$, $c_1$, and $c_2$) which keep the same sign of the discriminant of (\ref{polynomial}).
It makes the total number of cases four with both possible $\gammad$.
These four cases are presented in different panels of Fig. \ref{fig_distr} --
$\gammad=+1,\,\{c_0, c_1, c_2\} = \{1, 1, 1\}$ (discriminant of (\ref{polynomial}) is negative) in (a) panel, $\gammad=+1,\,\{c_0, c_1, c_2\} = \{-1, 1, 1\}$
(discriminant of (\ref{polynomial}) is positive) in (b) panel, $\gammad=-1,\,\{c_0, c_1, c_2\} = \{1, 1, 1\}$ in (c)
panel, and $\gammad=-1,\,\{c_0, c_1, c_2\} = \{-1, 1, 1\}$ in (d) panel. One can see that panels (a) and (c) demonstrate all possible roots combinations -- 
no roots (bottom region), two roots (middle one) and four roots (upper region),
while panels (b) and (d) demonstrate quite different picture. The (c) case have no roots region, but it is compact, while (d) case does not have it at all.

\begin{figure}
\includegraphics[width=1.0\textwidth, angle=0]{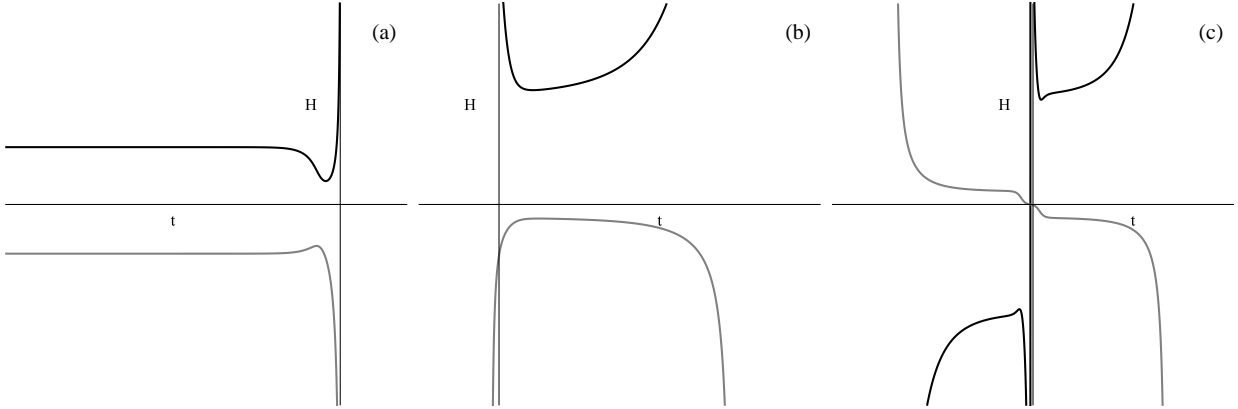}
\caption{Regimes that are possible in $\gammad=+1$ case with $\{c_0, c_1, c_2\} = \{1, 1, 1\}$ (see text for details).}\label{fig_k1_111}
\end{figure}

After describing the distribution of the number of roots over the parameter space we are going to describe all possible regimes in the considered model. On the following figures
black curves correspond to $H_a = H(t)$ variable while grey curves -- to
$H_b = \dot b(t)/b(t)$ -- we decided that plotting the variables with the same dimensionality is better for demonstration.

In Fig. \ref{fig_k1_111} we presented all regimes that are possible in $\gammad=+1$ case with $\{c_0, c_1, c_2\} = \{1, 1, 1\}$. If there are two roots, both branches are like
presented in Fig. \ref{fig_k1_111}(a); if there are four branches, one of them is like Fig. \ref{fig_k1_111}(a), two like Fig. \ref{fig_k1_111}(b), and the final is like
Fig. \ref{fig_k1_111}(c). One can see that all three of them have finite time future singularity, which makes it impossible to describe our Universe. The only
one of them which could compete for it, is Fig. \ref{fig_k1_111}(b) regime -- with increase of $b_0$ (the ``size'' of curved extra dimension) its ``lifetime'' increases, so with
large enough $b_0$ we can achieve long enough lifetime. But in turn it poses several problems, like how comes that initially the size of the extra dimension is that large, or
it could even remains large enough to be detected nowadays.

\begin{figure}
\includegraphics[width=0.75\textwidth, angle=0]{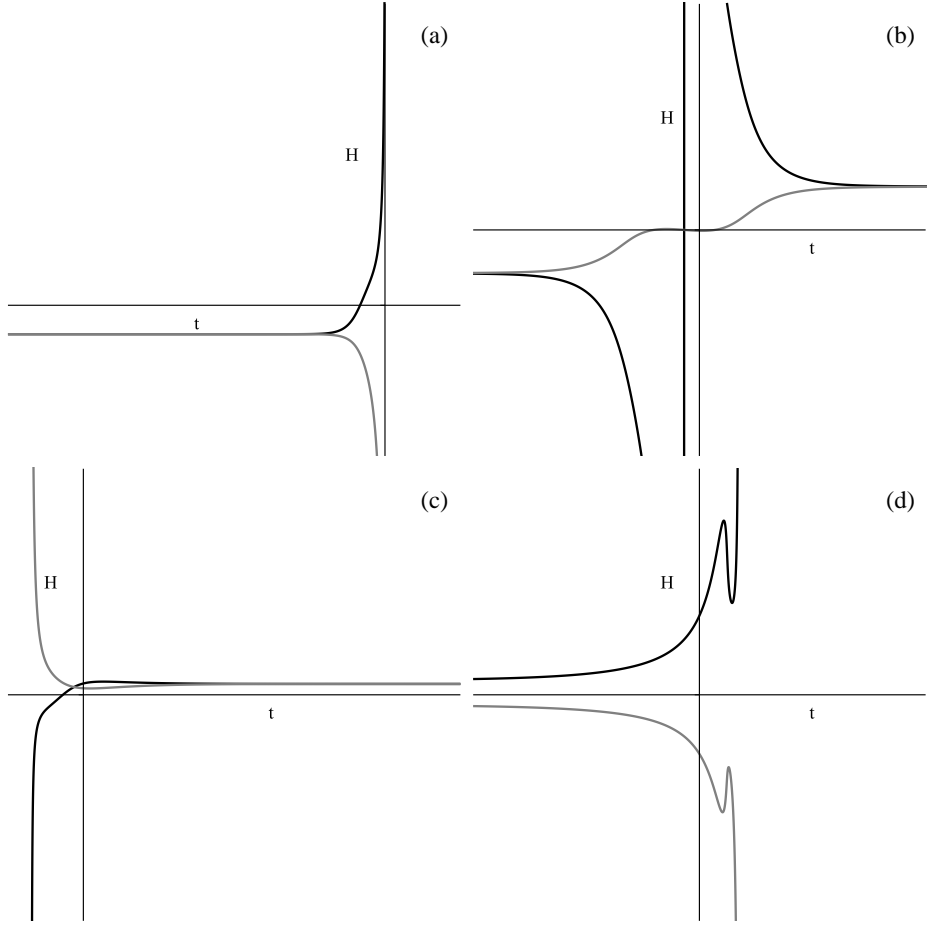}
\caption{Additional regimes that are possible in $\gammad=+1$ case (see text for details).}\label{fig_k1_-111}
\end{figure}

In Fig. \ref{fig_k1_-111} we presented all additional regimes that appear in $\gammad=+1$ case. If there are two roots in $\gammad=+1$ case with $\{c_0, c_1, c_2\} =
\{-1, -1, 1\}$, the regimes are like  in Fig. \ref{fig_k1_-111}(a) and Fig. \ref{fig_k1_-111}(b); if there are four roots in that case, the regimes are: one like Fig. \ref{fig_k1_-111}(a),
two like Fig. \ref{fig_k1_111}(b), and the final one is like Fig. \ref{fig_k1_-111}(b). Regime in Fig. \ref{fig_k1_-111}(c) is replacing the Fig. \ref{fig_k1_-111}(b) in the described
above picture in the $\{c_0, c_1, c_2\} = \{-1, 1, 1\}$ case. All the remaining cases with different nonzero $c_i$
are one of the already described three cases according to the sign of the discriminant of the (\ref{polynomial}).
Finally in Fig. \ref{fig_k1_-111}(d) we presented a regime that appears in $c_0 = 0$ case, where it
replaces Fig. \ref{fig_k1_111}(a) regime. The case with $c_1=0$ does not bring new regimes while $c_2= 0$ reduces the system to the Friedman dynamics.

At this stage we want to stress the attention of the reader to several points. Comparing Figs. \ref{fig_k1_-111}(a), (b), and (c) one cannot miss their familiarity. Indeed,
Fig. \ref{fig_k1_-111}(a) looks like left side of the Fig. \ref{fig_k1_-111}(b), while Fig. \ref{fig_k1_-111}(c) -- like its right side. But there is a difference between
Figs. \ref{fig_k1_-111}(a) and (c) from one side and Fig. \ref{fig_k1_-111}(b) from another -- in the former case the singularity is standard, while in the latter it is what
is called ``nonstandard singularity''. This kind of singularity is ``weak'' by Tipler's classification~\cite{Tipler}, and ``type II''
by Kitaura and Wheeler~\cite{KW1, KW2}. Recent studies of the singularities of this kind in the cosmological context in Lovelock and Einstein-Gauss-Bonnet gravity
demonstrates~\cite{mpla09, grg10, gc10, prd10} that the presence of this singularity is not suppressed and it is abundant for a wide range of initial conditions. And in our case
the singularities could ``turn'' one into another like in the just described above example.
So one can see that in our case the presence of this nonstandard singularity is also not suppressed; one can see another example of it in Fig. \ref{fig_k1_111}(c);
we discuss this type of singularity a bit additionally in Discussion.

Another point we want to stress an attention is the isotropisation and ``antiisotropisation'' issue. Comparing Fig. \ref{fig_k1_111}(a) with Fig. \ref{fig_k1_-111}(a) one can
see clearly that in former case we have  $H_b = - H_a$ in the past asymptote while in the latter -- it is $H_b = H_a$. The cause for that is the similar as in the classical
cosmology -- the presence of some isotropical matter, in our case it is an effective cosmological term (\ref{Leff}). In Fig. \ref{fig_k1_111}(a) case we have $\{c_0, c_1, c_2\} =
\{1, 1, 1\}$ so that $\Lambda_{eff}$ is imaginary and does not contribute to the dynamics -- we have ``antiisotropisation''. In contrast, in Fig. \ref{fig_k1_-111}(a) we have  $\{c_0, c_1, c_2\} =
\{-1, 1, 1\}$ which gives us real $\Lambda_{eff}$ so that we have isotropisation.  So that the $\Lambda_{eff}$ determines
both the roots distribution and the regimes.

\begin{figure}
\includegraphics[width=1.0\textwidth, angle=0]{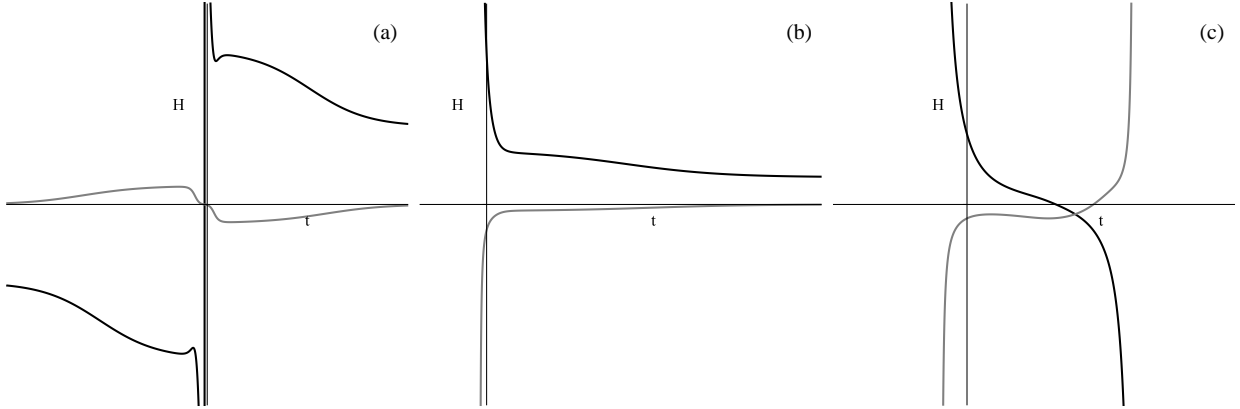}
\caption{Additional regimes that appear in the  $\gammad=-1$ case (see text for details).}\label{fig_k-1}
\end{figure}

Overall, the cases with $\gammad=+1$ do not give us ``well-behaved'' regimes because of the presence of finite time future singularities or because of isotropization, so we continue with $\gammad=-1$ case. Generally, most of the regimes remains in $\gammad=-1$ case,
but some new regimes added; we present them in Fig. \ref{fig_k-1}. From Fig. \ref{fig_k-1} one can immediately see that in (a) and (b) panels there are regimes without
finite time future singularity and with ``good'' behavior -- $H_a > 0$ and $H_b < 0$. The asymptotic behavior is like $H_a \to \const$, $H_b \to 0$; the latter imply $b(t) \to b_c$.
This is very attractive regime, and it occurs only in $\gammad=-1$ case.
The difference between Fig. \ref{fig_k-1}(a) and Fig. \ref{fig_k-1}(b) is clear -- in the former case
we have nonstandard singularity while in the latter we do not. The regime in the Fig. \ref{fig_k-1}(c) could be considered as a counterpart for Fig. \ref{fig_k1_111}(c) but with
no nonstandard singularity. The desired regime (in Fig. \ref{fig_k-1}(b)) is one of the four roots in the $\gammad=-1,\,\{c_0, c_1, c_2\} = \{1, 1, 1\}$ case and in the
similar cases (that with imaginary $\Lambda_{eff}$ in the sense of (\ref{Leff})). Therefore all possible regimes the only physically reasonable one is when the theory does not admit a maximally symmetric vacuum and when the curvature of the extra dimensions is negative.

Up to here  we considered only $\gamma_{(3)}=0$, but afterwards we confirmed the same results with both negative and positive $\gamma_{(3)}$, as the differences were only minor. More precisely, the
$\gamma_{(3)}>0$ case has only Fig. \ref{fig_k-1}(a) behavior while $\gamma_{(3)}<0$ has Fig. \ref{fig_k-1}(b)  as well. Also worth to mention, we have not detected numerically theoretically expected
regime with both $a(t)\to \const$ and $b(t)\to \const$ -- probably due to zero measure of the initial conditions which lead to it.

%To the best of authors knowledge this is the first time that a dynamical compactification in which the size of extra dimensions approaches to a constant while the three macroscopic extra dimensions
%expand has been achieved without fine tunings or ad hoc matter fields in any number of extra dimensions. Similar behavior was achieved in \cite{add13}, but the nature

\section{Discussion and comments}

\label{Discussion}

In this paper we have investigated numerically the dynamical compactification for the most generic Einstein-Gauss-Bonnet gravity in arbitrary dimensions. 
Explicit time-dependence of the scale factors of both 3-space and extra dimensions is assumed.
%To the best knowledge of the authors it is the first time where the explicit time dependence has been considered for both the expansion factor of the compact
%dimensions and for the expansion factor of the the three dimensional space. 
Moreover the constant curvatures of the space and compact dimensions were allowed
to be independent and to have any sign. Perhaps the case of an extra dimensional manifold with negative constant curvature is less familiar. It is however 
important to stress that a manifold of negative constant curvature can be compactified by taking the quotient by a freely acting discrete subgroup so that 
it will have a finite volume \cite{MalMao}. As the geometrically most simple compact manifold is a constant curvature one most simple there is no good 
a priori reason discarding  a compact space with negative constant curvature. It is also important to stress that there is no problem to define gravity if 
the extra dimensions have negative constant curvature as the behavior of the graviton depends on the effective Newton constant defined from 
(\ref{effectivenewton}) as

\begin{equation}
\begin{array}{l}
G_{N, eff} = (c_1(D+1)+2c_2(D-1)\Lambda_D);
\end{array}\label{eq_my_1}
\end{equation}

\noindent from it one can see that the effective Newton constant depends only on coupling constant and not on curvature and through it negative constant 
curvature do not prevent effective (3+1)-dimensional gravity from being well-defined.

One of the important results of this paper is that the different dynamical regimes must be classified according to the value of the effective cosmological 
constant (\ref{Leff}) rather then to the value of the ``lambda term''.

It was found that most of the regimes do not lead to physically reasonable scenarios as they develop either finite future time singularities (standard
or nonstandard) or isotropization of the entire space-time, where the size of the extra dimensions becomes of the same size than the one of the macroscopic 
three dimensional ones. It is worth pointing out that the singularities found in our analysis are quite different from the more common ones such as Big Rip
and similar to them (see e.g. \cite{McInnes, CKW, ENOW, NOT}) -- indeed, in our case the singularities are anisotropic so that the volume of the entire space do not need to diverge -- it could
become singular or even remain constant -- the latter is true for nonstandard singularities. With regard to the nonstandard singularities it is also
worth mentioning that the same kind of singularity but called ``determinant singularity'' was found for Bianchi-I with dilaton in~\cite{add14}.
This proves that the nature of this ``nonstandard singularity'' lies entirely in nonlinearity of the equations of motion.
%It is worth pointing out that the singularities found in our analysis are quite different from the more common big rip or singularities as 
%discussed  for example in \cite{McInnes}, \cite{CKW}, \cite{ENOW}, \cite{NOT}. Indeed in this case the singularities are anisotropic and it can happen that 
%one expansion factor diverges the other shrinks to zero at the same time so that the volume remains constant.

The only phenomenologically sensible scenario turned out to be the the case when the curvature of the extra dimensions is negative and when one chooses the
region of the couplings space $\{c_0,\,c_1,\,c_2\}$  so that  the theory does not admit a maximally symmetric vacuum solution (or in other words when the 
effective cosmological constant is complex). This scenario, which is independent from the sign of $\gamma_{(3)}$, can  be interpreted as a symmetry breaking 
mechanism. Remarkably it does not require neither fine-tunings, as the region of the parameter space is an open set, nor violations of ``naturalness 
hypothesis''. As we mentioned in the Introduction, the similar result -- the dynamical compactification without violation of ``naturalness'' -- was proposed
in~\cite{add13} for the (5+1)-dimensional
EGB theory. As one can see, our setup is different from the one used in \cite{add13} -- we considered both $M_D$ and $M_4$ as manifolds with constant and
possibly nonzero curvature, which gives a rise
to additional curvature terms and, as a consequence, to a new regime. Additionally, we considered all possible geometrical terms, including the volume term
($c_0\ne 0$), which crucially affects
the dynamics. Overall, despite the fact that in both cases -- in \cite{add13} and in our paper -- we can see dynamical compactification, it is brought by
different phenomena. In our case it is
``geometric frustration'' (see~\cite{CGP1}), which is brought by a combination of nonzero curvature and nonzero volume term -- both of them are usually
omitted from consideration since they complicate the equations a lot. The presence of non-zero curvature in the four dimensional part of the metric does
not change qualitatively the the analysis.

In the analysis we have supposed  the torsion to be zero. However in first order formalism the equations of motion of EGB gravity do not imply the vanishing 
of torsion which is therefore a propagating degree of freedom \cite{TZ-CQG}. Indeed exact solutions with non-trivial torsion have been found 
\cite{CGT07, CGW07, CG, CG2, ACGO}.  To study its effects in the context of dynamical compactification will be object of future investigation.

EGB gravity is the simplest generalization of General Relativity within the class of Lovelock gravity. As we considered an arbitrary number of extra 
dimensions it would also be interesting to study the effect of higher order Lovelock terms in the compactification mechanism. The results obtained in this 
paper suggest that the effects of higher terms depend sensibly on the fact if the highest curvature power is even as only in this case there exist a region 
in the parameter space which admits no maximally symmetric solution.

In the context of EGB theory existence of this region is solely due to the $c_0\ne 0$, as one can see from (\ref{Leff}). Taking into consideration this 
($c_0\ne 0$) case obviously makes the system far more difficult to solve -- especially if one look for exact solutions -- but it bears its fruits, as one 
can see. Indeed, when looking for exact solutions often only the simplest case is considered, and additional solutions are lost; this might be the case why 
this behavior has not been found earlier.

\bigskip

\textit{Acknowledgments.-- }
This work
was supported by FONDECYT grants 1120352, 1110167, and 3130599. The Centro de Estudios Cientificos (CECs) is funded
by the Chilean Government through the Centers of Excellence
Base Financing Program of Conicyt. We are also grateful for referees for their comments which help us improve the manuscript.


\begin{thebibliography}{999}


\bibitem{Lovelock} D.~Lovelock, J. Math. Phys. \textbf{12}, 498 (1971).
\bibitem{GastGarr} C. Garraffo and G. Giribet, Mod. Phys. Lett. {\bf A23}, 1801 (2008).

\bibitem{KK1} T. Kaluza, Sit. Preuss. Akad. Wiss. {\bf K1}, 966 (1921).
\bibitem{KK2} O. Klein, Z. Phys. {\bf 37}, 895 (1926).
\bibitem{KK3} O. Klein, Nature {\bf 118}, 516 (1926).
\bibitem{CGTW} F.~Canfora, A.~Giacomini, R.~Troncoso, and S.~Willison,
Phys. Rev. {\bf D80}, 044029 (2009)  [arXiv:0812.4311 [hep-th]].
%%\bibitem{BD85} D. G. Boulware and S. Deser, Phys. Rev. Lett. \textbf{55}, 2656 (1985).
\bibitem{add_1} F. M${\ddot {\rm u}}$ller-Hoissen, Phys. Lett. {\bf 163B}, 106 (1985).
\bibitem{add_2} J. Madore, Phys. Lett. {\bf 111A}, 283 (1985).
\bibitem{add_3} J. Madore, Class. Quant. Grav. {\bf 3}, 361 (1986).
\bibitem{add_4} F. M${\ddot {\rm u}}$ller-Hoissen, Class. Quant. Grav. {\bf 3}, 665 (1986).
\bibitem{Is86} H. Ishihara, Phys. Lett. \textbf{B179}, 217 (1986).
\bibitem{add_5} N. Deruelle, Nucl. Phys. {\bf B327}, 253 (1989).
\bibitem{44} J. Kripfganz and H. Perlt, Acta Phys. Polon. B {\bf 18}, 997 (1987).

\bibitem{add_6} T. Verwimp, Class. Quant. Grav. {\bf 6}, 1655 (1989).
\bibitem{add_9} N. Deruelle and L. Fari${\tilde {\rm n}}$a-Busto, Phys. Rev. D {\bf 41}, 3696 (1990).
\bibitem{add_10} J. Demaret, H. Caprasse, A. Moussiaux, P. Tombal, and D. Papadopoulos,  Phys. Rev. D {\bf 41}, 1163 (1990).
\bibitem{add_7} G. A. Mena Marug\'an, Phys. Rev. D {\bf 42}, 2607 (1990).
\bibitem{add_8} G. A. Mena Marug\'an, Phys. Rev. D {\bf 46}, 4340 (1992).


\bibitem{addn_1} T. Torii and H. Maeda, Phys. Rev. D {\bf 71}, 124002 (2005).
\bibitem{addn_2} T. Torii and H. Maeda, Phys. Rev. D {\bf 72}, 064007 (2005).
\bibitem{addn_3} J. Grain, A. Barrau, and P. Kanti, Phys. Rev. D {\bf 72}, 104016 (2005).
\bibitem{addn_4} R. Cai and N. Ohta, Phys. Rev. D {\bf 74}, 064001 (2006).
\bibitem{addn_5} H. Maeda, Phys. Rev. D {\bf 73}, 104004 (2006).
\bibitem{addn_6} M. Nozawa and H. Maeda, Class. Quant. Grav. {\bf 23}, 1779 (2006).
\bibitem{addn_7} H. Maeda, Class. Quant. Grav. {\bf 23}, 2155 (2006).
\bibitem{addn_8} M. Dehghani and N. Farhangkhah, Phys. Rev. D {\bf 78}, 064015 (2008).

\bibitem{add13} E. Elizalde, A.N. Makarenko, V.V. Obukhov, K.E. Osetrin, and A.E. Filippov,  Phys. Lett. {\bf B644}, 1 (2007).

%\bibitem{add_11} M. Farhoudi, Gen. Rel. Grav. {\bf 41}, 117 (2009).
\bibitem{add_12} A.~Toporensky and P.~Tretyakov, Gravitation \& Cosmology {\bf 13}, 207 (2007).

\bibitem{mpla09} S.A. Pavluchenko and A.V. Toporensky, Mod. Phys. Lett. {\bf A24}, 513 (2009).

\bibitem{grg10}   I.V. Kirnos, A.N. Makarenko, S.A. Pavluchenko, and A.V. Toporensky, Gen. Rel. Grav. {\bf 42}, 2633 (2010).
\bibitem{gc10} I.V. Kirnos, S.A. Pavluchenko, and A.V. Toporensky, Gravitation \& Cosmology {\bf 16}, 274 (2010).
\bibitem{prd10} S.A. Pavluchenko,   Phys. Rev. D {\bf 82}, 104021 (2010).
\bibitem{CGP1} F.~Canfora, A.~Giacomini and S.~A.~Pavluchenko,  Phys.\ Rev.\ D {\bf 88}, 064044 (2013)
\bibitem{MO04} K.I. Maeda and N. Ohta, Phys. Rev. D \textbf{71}, 063520 (2005).

\bibitem{MO14} K.I.~Maeda and N.~Ohta, JHEP {\bf 1406}, 095 (2014).

\bibitem{iv10-1} V.D. Ivashchuk, Int. J. Geom. Meth. Mod. Phys. {\bf 7}, 797 (2010).
\bibitem{iv10-2} V.D. Ivashchuk, Gravitation \& Cosmology {\bf 16}, 118 (2010).
\bibitem{prd09} S.A. Pavluchenko,   Phys. Rev. D {\bf 80}, 107501 (2009).

\bibitem{new12} S.A. Pavluchenko and A.V. Toporensky, Gravitation \& Cosmology {\bf 20}, 127 (2014).
\bibitem{new13} D. Chirkov, S. Pavluchenko, and A.V. Toporensky, Mod. Phys. Lett. {\bf A29}, 1450093 (2014).
% end of 22
\bibitem{new14} D. Chirkov, S.A. Pavluchenko, and A.V. Toporensky, arXiv:1403:4625


\bibitem{infl1} S.A. Pavluchenko, Phys. Rev. D {\bf 67}, 103518 (2003).

\bibitem{infl2} S.A. Pavluchenko, Phys. Rev. D {\bf 69}, 021301 (2004).



\bibitem{wolf} J.A. Wolf, {\it Spaces of constant curvature}, 4th edition (Publish or Perish, Wilmington, Delaware USA, 1984), p. 69.

\bibitem{Tipler} F.J. Tipler, Phys. Lett. {\bf A64}, 8 (1977).

\bibitem{KW1} T. Kitaura and J.T. Wheeler, Nucl. Phys. {\bf B355}, 250 (1991).

\bibitem{KW2} T. Kitaura and J.T. Wheeler, Phys. Rev. D {\bf 48}, 667 (1993).

\bibitem{MalMao} J. Maldacena and L. Maoz, JHEP {\bf 0402} 053 (2004).

\bibitem{McInnes} B. McInnes, JHEP {\bf 0208}, 29 (2002).
\bibitem{CKW} R.R. Caldwell, M. Kamionkowski, and N.N. Weinberg, Phys. Rev. Lett. {\bf 91}, 071301 (2003).
\bibitem{ENOW}E. Elizalde, S. Nojiri, S.D. Odintsov, and P. Wang, Phys. Rev. D {\bf 71}, 103504 (2005).
\bibitem{NOT} S. Nojiri, S.D. Odintsov, and S. Tsujikawa, Phys. Rev. D {\bf 71}, 063004 (2005).

\bibitem{add14} S. Alexeyev, A. Toporensky, and V. Ustiansky, Phys. Lett. {\bf B509}, 151 (2001).

\bibitem{TZ-CQG} R.~Troncoso and J.~Zanelli,
Class. Quant. Grav. \textbf{17}, 4451 (2000) [arXiv:hep-th/9907109].

\bibitem{CGT07} F.~Canfora, A.~Giacomini, and R.~Troncoso, Phys. Rev. D {\bf 77}, 024002 (2008).

\bibitem{CGW07} F.~Canfora, A.~Giacomini, and S.~Willison, Phys. Rev. D {\bf 76}, 044021 (2007).



\bibitem{CG} F.~Canfora and A.~Giacomini, Phys. Rev. D {\bf 78}, 084034 (2008).

\bibitem{CG2} F.~Canfora and A.~Giacomini, Phys. Rev. D {\bf 82}, 024022 (2010).

 \bibitem{ACGO}  A.~Anabalon, F. Canfora, A. Giacomini, J. Oliva,
 Phys. Rev. D {\bf 84}, 084015 (2011).


\end{thebibliography}
\end{document}